\documentclass[
   aps,
   prb,
   superscriptaddress,
   twocolumn,
   showpacs]{revtex4-1}

\usepackage{amsmath,amssymb,bm,epsfig}
\usepackage{dcolumn}
\usepackage{bm}
\usepackage{feynmf}
\usepackage{amsmath,amssymb,bm,epsfig}
\usepackage{dcolumn}
\usepackage{bm}

\usepackage{color}
\usepackage{siunitx}
\usepackage{graphicx}
\usepackage{amsthm}
\usepackage{amsmath}
\usepackage{amssymb}
\usepackage{enumerate}
\usepackage[table]{xcolor}
\usepackage{booktabs}
\usepackage{xcolor}

\AtBeginDocument{
\heavyrulewidth=.08em
\lightrulewidth=.05em
\cmidrulewidth=.03em
\belowrulesep=.65ex
\belowbottomsep=0pt
\aboverulesep=.4ex
\abovetopsep=0pt
\cmidrulesep=\doublerulesep
\cmidrulekern=.5em
\defaultaddspace=.5em
}

\begin{document}

\title{Anomalous Transition Magnetic Moments in two-dimensional Dirac Materials }

\author{Sanghita Sengupta}
   \affiliation{Institut Quantique and D$\acute{e}$partement de Physique, Universit$\acute{e}$ de Sherbrooke, Sherbrooke, Qu$\acute{e}$bec, Canada J1K 2R1}
   \author{Madalina I. Furis}
   \affiliation{Materials Science Program, University of  Vermont, Burlington, VT 05405}
   \affiliation{Department of Physics, University of  Vermont, Burlington, VT 05405}
   \author{Oleg P. Sushkov}
   \affiliation{School of Physics, University of New South Wales, Sydney 2052, Australia}
   \author{Valeri N. Kotov}
   \affiliation{Department of Physics, University of  Vermont, Burlington, VT 05405}
   \affiliation{Materials Science Program, University of  Vermont, Burlington, VT 05405}
\date{\today}
\begin{abstract}
     We show that the magnetic response of atomically thin materials with Dirac spectrum and spin-orbit interactions can exhibit  strong dependence on electron-electron interactions. While graphene itself has a very small spin-orbit coupling, various two-dimensional (2D) compounds ``beyond graphene'' are good candidates to exhibit the strong interplay between spin-orbit and Coulomb interactions. Materials in this class include dichalcogenides (such as MoS$_2$ and WSe$_2$), silicene, germanene, as well as 2D topological insulators described by the Kane-Mele model. We present a unified theory for their in-plane magnetic field response leading to ``anomalous'', i.e. electron interaction dependent transition moments. Our predictions can be potentially used to construct unique magnetic probes with high sensitivity to electron correlations.
\end{abstract}
\maketitle

\section{Introduction}
Two-dimensional quantum materials are characterized by low-energy quasiparticle excitations that can be fully described by an effective (2+1)-dimensional Dirac equation. Naturally, various quantum electrodynamics (QED) phenomena associated with Dirac physics manifest themselves in these quantum condensed matter systems\cite{Neto2009, Katsnelson2006,KATSNELSON20073, Kotov} even though the Dirac quasiparticles have non-relativistic nature and arise purely from band structure considerations.

One such astonishing feature associated with this class of materials is their magnetic response. In the presence of a magnetic field, the Dirac fermions exhibit a plethora of quantum phases which can range from anomalous quantum Hall states \cite{Novoselov2005197,Zhang2005} to quantum holography in graphene flakes\cite{franz}. While most studies related to anomalous quantum Hall physics have been conducted within the context of massless 2D Dirac fermions\cite{gusynin,zheng,Zhang2005,Novoselov2005197}, recent research elucidates similar magnetic phenomena arising in the regime of massive 2D Dirac fermions \cite{offidani, murakami}. 

In this paper we explore the magnetic response of the massive 2D Dirac fermions, with a special focus on the effect of electron-electron interactions. The candidate materials for this study include:  (i) quantum Spin Hall  (QSH)  insulator states described by the Kane-Mele model\cite{kane}, (ii) atomically thin semiconductor family of transition metal dichalocogenides (TMDCs)\cite{Novoselov10451,xiao} and (iii) topological insulator family of Silicene-Germanene class of materials\cite{cahangirov,vogt}. 

Our chosen materials are characterized by  gapped Dirac spectrum. In the case of QSH states described by the Kane Mele model, it was shown that the symmetry allowed spin-orbit coupling (SOC) leads to an opening of the energy gap in the linear, gapless electronic dispersion of graphene\cite{kane}. This SOC thus converts the 2D semi-metallic graphene into a 2D topological insulator with gapless edge modes, while being insulating in the bulk. These QSH states thus allow for the generation of dissipationless spin currents and are a topic of immense interest\cite{kane,hasankane,kane2005}. However, it was also pointed out that while the SOC in graphene is of the order of 4 meV, the gap generated by it is rather small, of the order of $\sim$ 10$^{-3}$ meV\cite{yao,macdonald}.
One of the goals of the present work is to study in detail the in-plane magnetic response of the Kane Mele model where we show that Coulomb interactions can have quite significant effect and lead to enhanced spin flip (transition) magnetic moment.

Our theoretical approach is conceptually similar to calculations performed in relativistic QED \cite{landaulifshitz4,schwinger} where 
 Schwinger's celebrated vertex correction to the Dirac  electron magnetic form factor  translates into anomalous (fine structure constant dependent) g-factor.
 Of course all materials considered in this work are non-relativistic systems with effective Dirac quasiparticles;
  thus  any ``anomalous" corrections to spin response 
  will originate from the Coulomb interaction between quasiparticles.
 Naturally, the results for the Kane Mele model and the other  2D systems with SOC  will be anisotropic  since it is well known that all of them exhibit strong intrinsic spin anisotropy, with the  spin z-component (perpendicular to the planes) conserved.
  This means that only $\textit{in-plane}$ magnetic fields, leading to off-diagonal (spin flip) transitions, can give rise to anomalous, i.e. Coulomb interaction dependent $\textit{transition}$ magnetic moments. We also point out that interaction-dependent magnetic moments have recently been studied for three dimensional Dirac and Weyl insulators \cite{stoof}. Compared to those systems, the spin response of 2D materials with SOC is also, naturally, quite different and we describe it in detail  in this work.

As mentioned before, we will also extend and apply our formalism and calculations of  anomalous transition magnetic moments to two other systems which include the atomically thin TMDCs and Silicene-Germanene class of materials. Besides being gapped, these materials also display strong intrinsic spin-orbit coupling effects\cite{hatami,xiao, ezawa1,tabert1,tabert2,tabert3,tabert4,tabert5,tabert6}.  Thus the  interplay of electron-electron interactions and SOC in these systems is a topic of great interest.

The general structure of the paper is as follows: we will begin with the Kane Mele model in Sec.~\ref{sec:KM}, providing the general methodology and results for the one-loop correction to the transition magnetic moment. We will then adapt and extend this formalism to TMDCs and Silicene-Germanene class of materials in Sec.~\ref{sec:TMDC} and Sec.~\ref{sec:SG}. Finally we will conclude in Sec.~\ref{sec:sum} with an outlook that summarizes our results. We also discuss possible experimental probes for detection of the interaction effects calculated  in this work.

\section{Effect of Coulomb interactions on transition (spin-flip) magnetic moment within the Kane Mele model}
\label{sec:KM}
The Kane Mele model describes the general 2D Dirac Hamiltonian with a mass term that originates from the spin-orbit coupling. This SOC renders the system as gapped and much of this section will be devoted to understanding the interplay of Coulomb interactions and the SOC in relation to transverse magnetic response. Let us begin with the general procedure to calculate the one-loop correction to transition moment for this model. 
The Hamiltonian of the Kane Mele model\cite{kane} is 
\begin{equation}\label{hamKM}
H =  v{\bf\sigma} \cdot {\bf k} + \lambda \sigma_z s_z ,
\end{equation}
where  $v$ is the Fermi velocity in the material. It is convenient, and customary in the literature, to
 label the  spin $z$ component  of the fermion as lowercase $s_{z}=\pm 1$, for spins up and down. The Pauli matrices $\hat{\sigma_{i}}$  act in pseudospin (sublattice) space and   the spin-orbit coupling is given by $\lambda$. In our derivations we choose the convenient natural units $\hbar =v =1$, unless otherwise mentioned. From the Hamiltonian we can see that the spin in the ${z}$ channel is always conserved. This means that  interaction corrections to the diagonal (same spin) transitions are forbidden, while spin-flip transitions (caused by  magnetic field in the $S_{x}$ (or $S_{y}$) direction) can acquire  Coulomb interaction - dependent  components. 
 We refer to such interaction contributions as ``anomalous" spin response components. 

Without loss of generality,  in this and the next sections, we work in a given valley (already assumed in the above Hamiltonian). It is easy to see that the results for the spin response are valley independent (which also applies to the interaction corrections since 
the long-range Coulomb interaction does not mix valleys.) We will also be assuming, in this and all other sections, that the system always remains an insulator (i.e. the chemical potential is in the gap).
 
Within the Hamiltonian of the Kane Mele model, the dispersion relation $\varepsilon_{k}$ and eigenfunctions at momentum $k$ are given as
\begin{equation}\label{eigenKM}
\varepsilon_{k} = \pm\sqrt{k^2 + \lambda^2} ,
\end{equation}
\begin{equation}\label{cupKM}
\Psi(k)_{+} = \frac{k}{\sqrt{2}\sqrt{\varepsilon_k^{2}-\lambda |\varepsilon_k|}}
\left( \begin{array}{c} 1 \\
\frac{|\varepsilon_k|-\lambda}{(k_x-ik_y)} 
\end{array} 
\right),
\end{equation}
\begin{equation}\label{cdownKM}
\Psi(k)_{-} = \frac{k}{\sqrt{2}\sqrt{\varepsilon_k^{2}+\lambda |\varepsilon_k|}}
\left( \begin{array}{c} 1 \\
\frac{|\varepsilon_k|+\lambda}{(k_x-ik_y)} 
\end{array}
\right).
\end{equation}
The wave functions 
$\Psi(k)_{s}$ are labeled by the spin index $s_z=s= \pm 1$.

Next, we consider coupling to a uniform in-plane magnetic field   of the form $B_{x}S_{x}$, with the coupling constant given by the $g$ factor times the effective Bohr magneton set to one for 
convenience, $g \mu_B =1$. We define a quantity we call bare  transition magnetic moment as $\mu =2 \langle\downarrow|S_{x}|\uparrow\rangle$. 
Here  the (normalized)  spin up state is a product of the pseudospin and spin  wave functions: $|\uparrow\rangle = \Psi(k)_{+} \chi_{+} $, where  $\chi_{+} = \left ( \begin{array}{c} 1 \\  0 \end{array}\right )$ 
is the spin up spinor in  spin space. 
Similarly: $|\downarrow\rangle = \Psi(k)_{-} \chi_{-} $, $\chi_{-} = \left ( \begin{array}{c} 0 \\  1 \end{array}\right )$ . From the usual spin $1/2$ algebra we have: $2 S_x \chi_{+} = \chi_{-}$.

  Using the above wavefunctions, we calculate the bare transition moment for this model:
\begin{eqnarray}\label{bareKM}
\mu& =& 2 \langle \downarrow |  S_x | \uparrow \rangle =  (\chi_{-}^\dagger  (2 S_x) \chi_{+}) ( \Psi(k)_{-}^\dagger  \Psi(k)_{+}) \nonumber \\
&=&  \Psi(k)_{-}^\dagger  \Psi(k)_{+} = \frac{k}{\sqrt{k^{2}+\lambda^{2}}}. 
\end{eqnarray} 
From now on we will use the shorthand notation $|\uparrow\rangle, |\downarrow\rangle$ in all calculations in this section as well as for the models considered in
subsequent sections.

\begin{figure}[t!]
\begin{center}
\includegraphics[width=\columnwidth]{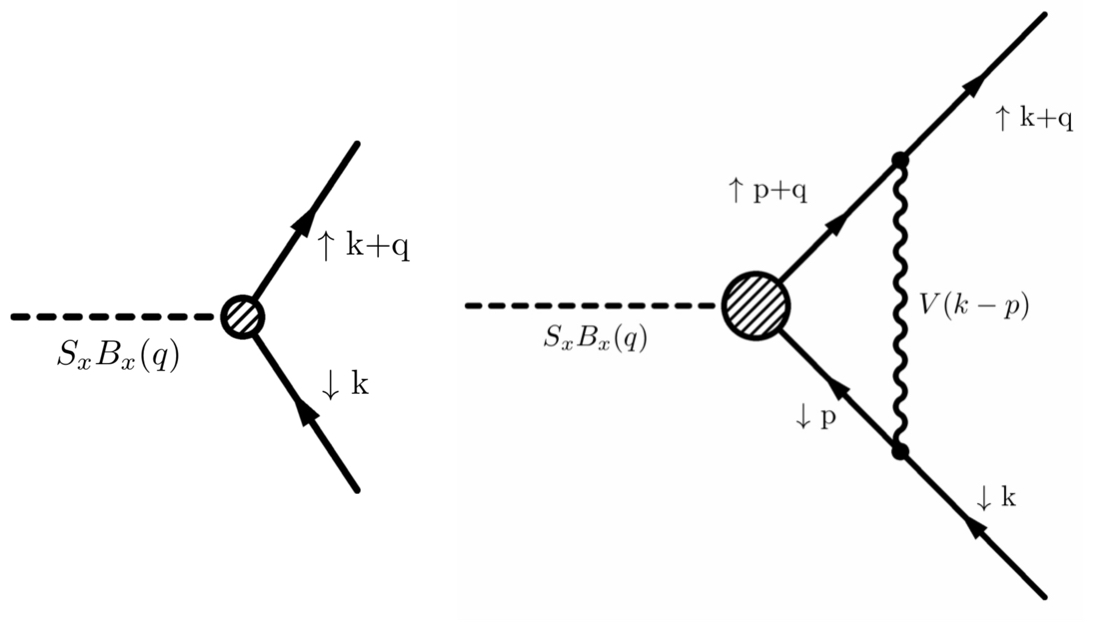} 
\caption{\label{fig:BFD} Left: Feynman diagram for the bare transition moment with uniform (zero momentum, $q \rightarrow 0$)  in-plane magnetic field  $B_{x}$, corresponding to 
field coupling of the form $B_{x}(q \rightarrow 0) S_{x}$.  We set the field coupling  prefactor $g \mu_{B} = 1$ in this field direction   for simplicity 
(it is known that the $g$-factor can be  strongly  material dependent and should be restored when comparison with experiment is made). 
 Right: Vertex diagram associated with the one-loop Coulomb interaction correction shown by the wiggly line V$({\bf p}) = 2\pi e^{2}/p$.}
\end{center}
\end{figure}

We proceed  to calculate the effect of electron-electron (Coulomb) interactions on the transition magnetic moment.
Basic Feynman diagrams for the bare and one-loop (vertex) correction are given in Fig.~\ref{fig:BFD}. Invoking Feynman rules we will write analytic expression corresponding to the vertex function given in the right panel of Fig.~\ref{fig:BFD}. 

Therefore the one-loop Coulomb interaction correction to the magnetic moment (for $q \rightarrow 0$) is given as
\begin{equation}\label{ICsiliKM}
\begin{split}
\delta\mu&=2 \sum_{p} i\int\frac{\mathrm{d}\omega}{2\pi}\langle\downarrow|G^{s=-1}(p,\omega)S_{x}G^{s=+1}(p+q,\omega)|\uparrow\rangle \\
&\quad \times V(|{\bf p}-{\bf k}|) \\
\end{split}
\end{equation}
where,  $V({\bf p}) = (2\pi e^{2}/p)$ is the Coulomb interaction and the corresponding Green's functions for this model is
\begin{equation}\label{GFKM}
G({\bf k},\omega) = \frac{\omega +({\bf \sigma} \cdot {\bf k} + \lambda \sigma_{z}s_{z})}{\omega^{2}-\varepsilon_{k}^{2} +i\eta}.
\end{equation}
Using the above equations along with the the corresponding wave functions (Eq.~(\ref{cupKM}) and Eq.~(\ref{cdownKM})), we derive an expression for the one-loop interaction correction,
\begin{equation}\label{delmKM}
\delta \mu = \frac{k}{\sqrt{k^{2}+\lambda^{2}}}\alpha \mathcal{W}(k/\lambda),
\end{equation}
where we have defined,
\begin{equation}\label{FKM}
\alpha \mathcal{W}(k/\lambda) = \frac{\lambda^2}{2} \int \frac{d^2p}{(2\pi)^2} \frac{V(|{\bf p}-{\bf k}|)}{|\varepsilon_{p}|^3} \left ( 1 - \frac{{\bf p}\cdot{\bf k}}{k^2}\right ),
\end{equation}
with $\alpha = e^{2}/\epsilon \hbar v$ as the effective fine-structure constant representing the strength of Coulomb interactions and $\epsilon$ is the dielectric constant.

\begin{figure}[h]
\begin{center}
\includegraphics[width=\columnwidth]{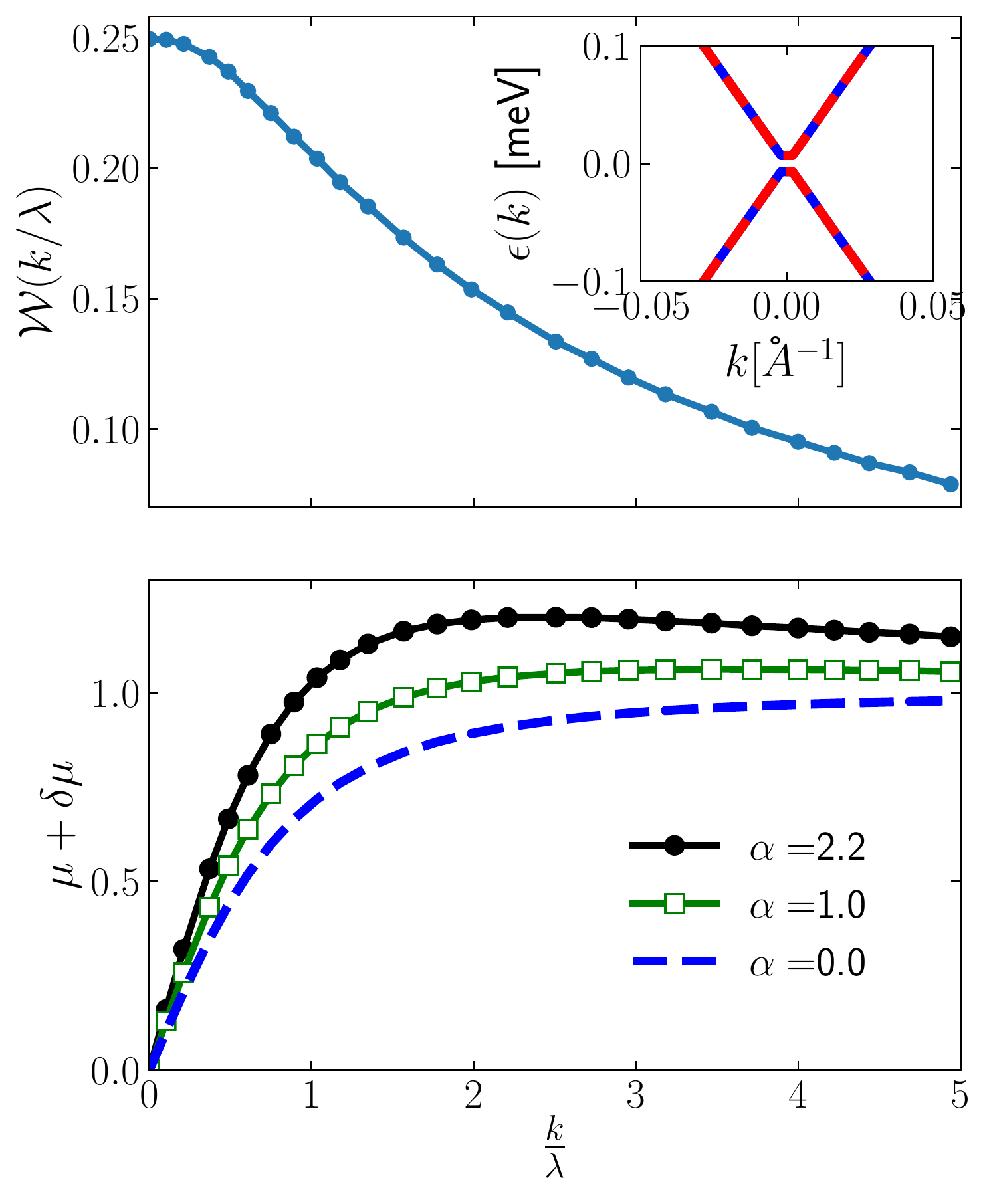} 
\caption{\label{fig:DCKM}Top panel: Variation of the  correction function $\mathcal{W}(k/\lambda)$ with rescaled momentum $k/\lambda$. The magnitude of the correction is large and maximum at $k =0$. Right inset: Low-energy band structure for the Kane Mele model. Bands are spin degenerate with a gap that is generated by the spin-orbit interaction $\lambda = 1 \mu$eV. Bottom panel: Variation of the total spin-flip transition moment, $\mu + \delta \mu = \frac{k}{\sqrt{k^2 + \lambda^2}}\left (1+  \alpha \mathcal{W}(k/\lambda) \right ) $,  with rescaled momentum ($k/\lambda$). With the increase in $\alpha$ we see an enhancement in the total transition moment. We relate this increase to the enlarged correction effects from Coulomb interactions.}
\end{center}
\end{figure}

The variation of the correction function $\mathcal{W}(k/\lambda)$ with the dimensionless band momenta ($k/\lambda$) is shown in the top panel of Fig.~\ref{fig:DCKM}. The Coulomb interaction correction  peaks at $k=0$ and there after decays with the increase in band momenta. 

Using Eqs.~(\ref{bareKM}), (\ref{delmKM}) and (\ref{FKM}), we write the total transition moment as,
\begin{equation}\label{TMKM}
\begin{split}
\mu + \delta \mu &=\frac{k}{\sqrt{k^2 + \lambda^2}}\bigg[ 1 + \frac{\lambda^2}{2} \int \frac{d^2p}{(2\pi)^2} \frac{V(|{\bf p}-{\bf k}|)}{|\varepsilon_{p}|^3}\\
&\quad \bigg( 1 - \frac{{\bf p}\cdot{\bf k}}{k^2} \bigg)
\bigg].
\end{split}
\end{equation}
To display the effects of the Coulomb interaction correction, we show the dependence of the total transition moment $\mu + \delta \mu$ with the dimensionless band momenta for various values of the coupling $\alpha$ in the bottom panel of Fig.~\ref{fig:DCKM}. 

The maximum value of $\alpha = 2.2$ can in principle be achieved in suspended samples, while 
 additional effects   leading to coupling constant renormalization due to self-consistent screening and/or substrate effects
 should also be taken into account. All of these lead to a decrease of the effective coupling. First, the presence of a  substrate 
with  dielectric constant   $\kappa$  will reduce the  Coulomb coupling $\alpha$ via: $\alpha \rightarrow \alpha/\epsilon$,
where $\epsilon = (1+\kappa)/2$, assuming the 2D material is on a substrate with air on the other side. For example the dielectric
constant of the commonly used SiO$_2$ is $\kappa \approx 4$, leading to a decrease of $\alpha$ by a factor of 2.5.
Second, due to the electron polarization  in the 2D material, the Coulomb interaction is screened, which can be taken
into account self-consistently within the usual RPA (random phase approximation) scheme. The effective Coulomb interaction is obtained 
by the simple replacement $V(k) \rightarrow V(k)/(1 - V(k)\Pi(k))$.  In this way  the results become reliable even in the regime of strong bare coupling (e.g. $\alpha = 2.2$).
 We present results for  static screening  which involves the static polarization function $\Pi(k,\omega=0)\equiv \Pi(k)$ for a material with a  gapped 2D Dirac spectrum\cite{Gorbar,vnk1}, appropriate for the Kane-Mele model:
\begin{equation}\label{pol}
\Pi(k) = -\frac{1}{\pi}\bigg(\frac{\lambda}{v^{2}}\bigg)-\frac{k}{2\pi v}\bigg[1-\frac{4\lambda^{2}}{v^{2}k^{2}}\bigg]\tan^{-1}\bigg(\frac{vk}{2\lambda}\bigg) .
\end{equation} 
When we incorporate the effects of the gapped polarization, the total transition magnetic moment transforms to
\begin{equation}\label{alpha}
\begin{split}
\mu + \delta \mu &=\frac{k}{\sqrt{k^2 + \lambda^2}}\bigg[ 1 + \frac{\lambda^2}{2} \int \frac{d^2p}{(2\pi)^2} \frac{V(|{\bf p}-{\bf k}|)}{|\varepsilon_{p}|^3}\\
&\quad \times \frac{1}{(1-V(|{\bf p}-{\bf k}|)\Pi(|{\bf p}-{\bf k}|))}\bigg( 1 - \frac{{\bf p}\cdot{\bf k}}{k^2} \bigg)
\bigg],
\end{split}
\end{equation}
where we have used again $v=1$.

\begin{figure}[h]
\begin{center}
\includegraphics[width=\columnwidth]{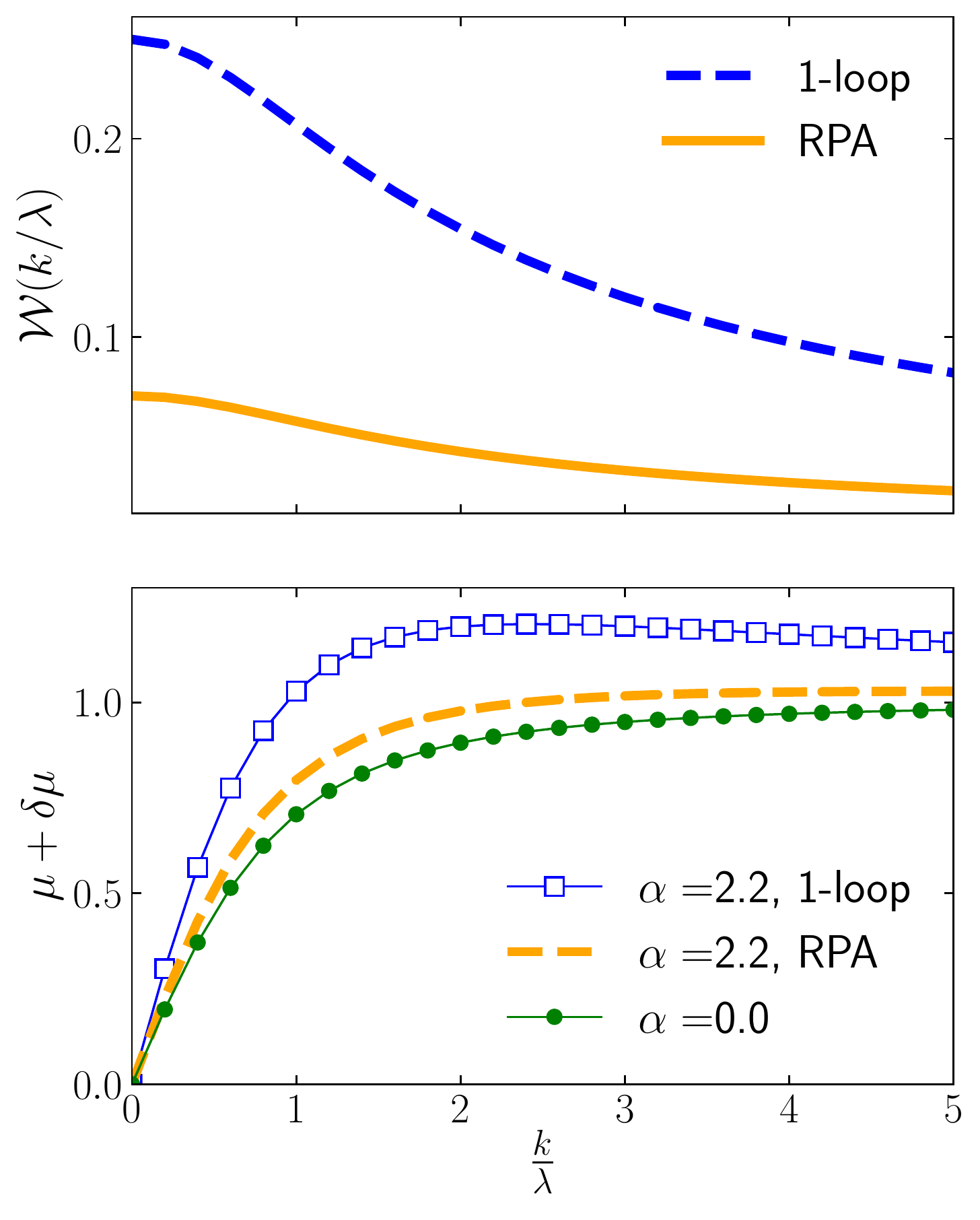} 
\caption{\label{fig:rpa}Top panel: A comparison of the correction function $\mathcal{W}(k/\lambda)$ with rescaled momentum $k/\lambda$ for the 1-loop case and the RPA. The magnitude of the correction is seen to decrease with the inclusion of self-consistent screening. Bottom panel: Comparative plots of the variation of the spin-flip transition moment, $\mu + \delta \mu $,  with rescaled momentum ($k/\lambda$) for the suspended case within the formalism of 1-loop, RPA and no-Coulomb correction effects. }
\end{center}
\end{figure}

Within RPA, assuming a suspended sample, the correction function as well as the total transition magnetic moment is shown in Fig.~\ref{fig:rpa}. It is evident that  self-consistent screening further decreases the correction function, as expected. This decrease is also manifested in the decrease of the total transition magnetic moment.

In our next section we extend this formalism to calculate the one-loop Coulomb interaction correction for the transition moment in the atomically thin dichalcogenides.


\section{Anomalous transition moment in atomically thin family of Dichalcogenides}
\label{sec:TMDC}
In contrast to the Kane Mele model, the atomically thin transition metal dichalcogenides (TMDCs) display a large spin-independent gap ($\sim$ of the order of few eVs) which originates from the broken inversion symmetry of the sublattice of these systems\cite{xiao,hatami}. Along with a large spin independent gap which we refer as $\Delta$, these materials also display strong intrinsic spin-orbit coupling arising from the admixture of the d-orbitals of the transition metals\cite{hatami}. In this section, we will probe the Coulomb interaction effect on the SOC-induced magnetic moment of these class of materials. Our procedure will be  the same as before.

\begin{figure}[h]
\begin{center}
\includegraphics[width=\columnwidth]{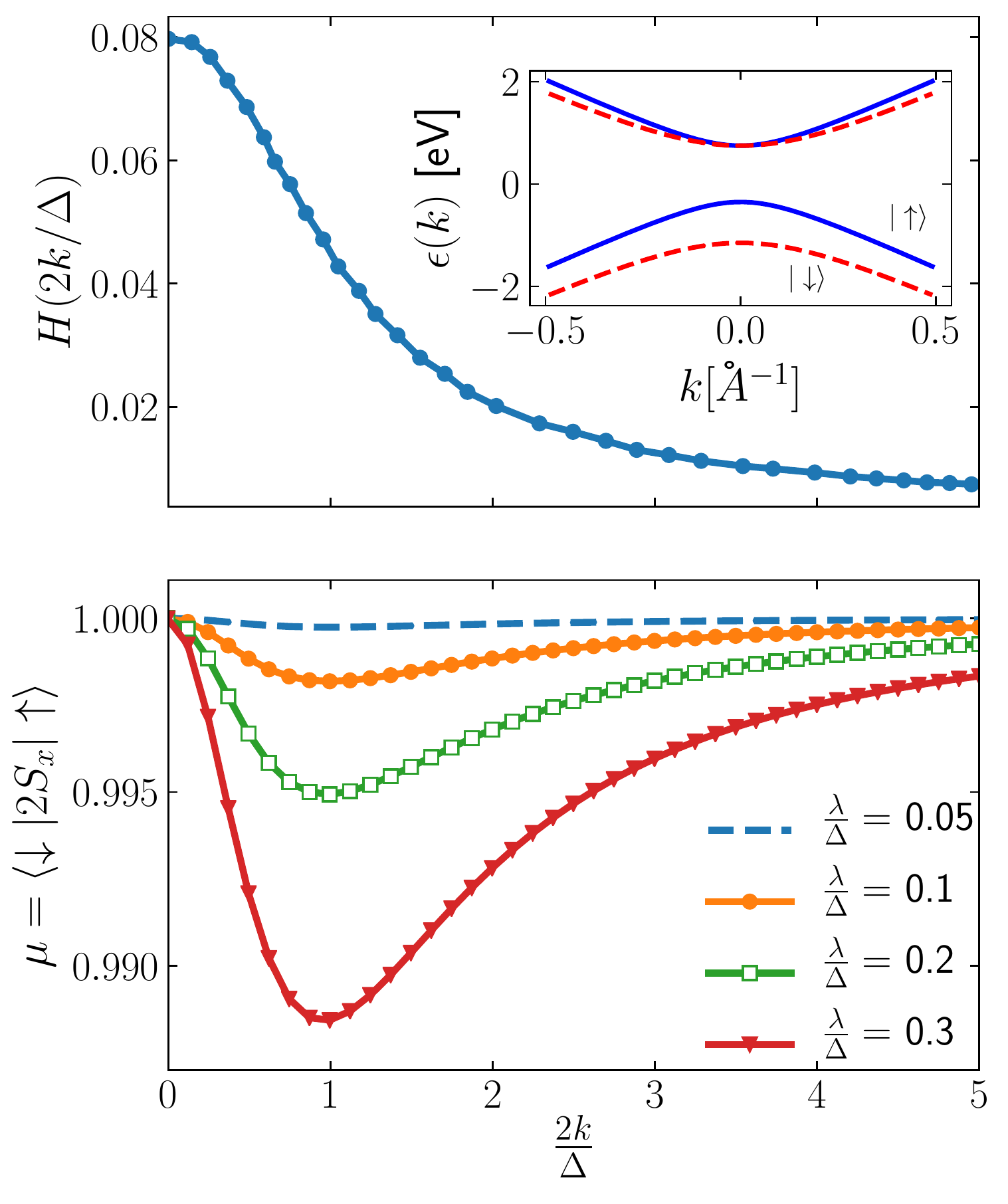} 
 
\caption{\label{fig:TMDC} Top panel: Correction function $H(2k/\lambda)$ for the dichalcogenides. Here, we have used the rescaled momentum: $2k/\Delta$. As can be seen from the figure, at $2k/\Delta =0$, the value of the correction function $H(2k/\Delta)$ is very small ($\sim 0.08$.) Inset plot on top panel shows the corresponding dispersion relation for the dichalcogenides. The conduction bands are degenerate at k$=0$ and are also seen to undergo a band inversion. Bottom panel: Bare transition moment for the dichalcogenides for various values of $\lambda/\Delta$. For TMDCs with the relevant value of $\lambda/\Delta \lesssim 0.15$, we see that the value of $\mu$ is almost a constant $\approx 1$ and shows negligible variation with the momentum.  }
\end{center}
\end{figure}

The effective low energy Hamiltonian associated with the monolayer TMDCs\cite{xiao},
\begin{equation}\label{hamTMDC}
H =  {\bf\sigma} \cdot {\bf k} + (\Delta/2) \sigma_z -(\lambda/2)(\sigma_z -1)s_z.
\end{equation}
Here, $\Delta$ is the spin-independent gap and $\lambda$ is the spin-orbit coupling. The model parameters for MoS$_2$  are $\Delta \approx 1.66$ eV, 2$\lambda \approx 0.15$ eV; 
for WS$_2$ are $\Delta \approx 1.79$ eV, 2$\lambda \approx 0.43$ eV, and for
WSe$_2$ are $\Delta \approx 1.6$ eV, 2$\lambda \approx 0.46$ eV \cite{xiao, hatami} which clearly indicate that the family of TMDCs can be classified by a regime in which the spin-independent gap is much larger compared to the spin-orbit coupling,
\begin{equation}
\Delta/\lambda \gg 1.
\end{equation}

The exact wave functions at momentum $k$ for these class of materials are written as
\begin{equation} \label{efTMDC}
\Psi(k)_n^{s} = \frac{k}{\sqrt{k^2 + (E_{k,n}^s)^2}}
\left( \begin{array}{c} 1 \\
E_{k,n}^s/(k_x-ik_y)
\end{array}
\right) , \  \  n=1,2
\end{equation}
with $s$ as the spin index, $n=1$ and $n=2$ labels the conduction band and valence band respectively. Here we have defined $E_{k,n}^s$ as the quantities
\begin{equation} \label{defTMDC}
E_{k,n}^s = \varepsilon_{k,n}^s - \Delta/2, \  \  \   s=s_z =\pm 1 .
\end{equation}
$\varepsilon_{k,n}^s$ represent the eigenenergies:
\begin{equation}\label{eeTMDC}
\varepsilon_{k,1}^s  = \lambda s/2  + \varepsilon_{k}^s \  > 0,  \  \ n=1,
\end{equation}
\begin{equation}
\varepsilon_{k,2}^s  = \lambda s/2  - \varepsilon_{k}^s  \ < 0,  \  \   n=2,
\end{equation}
with $\varepsilon_{k}^s$ appropriately defined:
\begin{equation}\label{defepsilonTMDC}
\varepsilon_{k}^s \equiv + \sqrt{k^2+[(\Delta -\lambda s)^2/4]}.
\end{equation}

In the right inset of top panel of Fig.~\ref{fig:TMDC}, we show the low-energy band structure for this group of materials corresponding to Eq.~(\ref{defepsilonTMDC}). These bands are non-degenerate, showing spin inversion with a large spin-independent gap. 

The bare transition moment is calculated using the wave functions for the conduction band (given in Eq.~(\ref{efTMDC})) leading to 
\begin{equation}\label{bareTMDC}
\mu = 2\langle \downarrow | S_x | \uparrow \rangle = \frac{k^2 + E_{k,1}^{+}E_{k,1}^{-}}{\sqrt{(k^2 + [E_{k,1}^{+}]^2)(k^2 + [E_{k,1}^{-}]^2)}}.
\end{equation}

For the correction to the bare transition moment we will use the vertex function and Eq.~(\ref{ICsiliKM}). The Green's function for this model is 
\begin{equation}\label{GFTMDC}
\begin{split}
    G ^{s}(p,\omega) &= \frac{1}{2 \varepsilon_{p}^s} \bigg[ \frac{\varepsilon_{p}^s + {\bf\sigma} \cdot {\bf p} + \sigma_z(\Delta -\lambda s)/2}{\omega -\varepsilon_{p,1}^s + i \eta}\\
    &\quad - \frac{-\varepsilon_{p}^s + {\bf\sigma} \cdot {\bf p} +\sigma_z (\Delta -\lambda s)/2}{\omega -\varepsilon_{p,2}^s - i \eta} \bigg] .\\
\end{split}
\end{equation}

Using the above Green's function we first perform the frequency integral in Eq.~(\ref{ICsiliKM}) with the result

\begin{equation}\label{FITMDC}
\begin{split}
    i \int \frac{d\omega}{2\pi}\bigg[G ^{-}G ^{+}\bigg] & \approx \frac{1}{4\varepsilon_{p}^{+}\varepsilon_{p}^{-}}\frac{1}{[\lambda^2 -(\varepsilon_{p}^{-}+\varepsilon_{p}^{+})^2]} \bigg[  
    \lambda^2 \frac{\Delta}{\varepsilon_{p}} ({\bf\sigma} \cdot {\bf p})\\
    &\quad -  2 \lambda^2 \frac{p^2}{\varepsilon_{p}} (\sigma_z + 1)
    - 4 \lambda \varepsilon_{p} ({\bf\sigma} \cdot {\bf p}) \sigma_z 
    \bigg].
\end{split}
\end{equation}
Here we have expanded the numerator up to O[$\lambda^{2}$]. The pre-factors of Eq.~(\ref{FITMDC}) given by the energy denominators can be taken at $\lambda =0$ because their expansion starts from a constant and the next order is O[$\lambda^{2}$]. Following Eq.~(\ref{ICsiliKM}), the interaction correction to the transition moment is derived by taking the expectation value of the above equation with respect to the wave functions $\Psi(k)_1^{\pm}$ (Eq.~(\ref{efTMDC})),
  
\begin{equation}\label{dmTMDC}
\delta \mu =
\sum_{{\bf p}} V(k-p) \ \frac{(-1)}{16 \varepsilon_{p}^4} \  \lambda^2  \ 
\frac{2}{k^2 + E_k^2} \ \Gamma(p,k),
\end{equation}
where the function $\Gamma$(p,k) has been calculated as
\begin{equation}\label{gammaTMDC}
\begin{split}
\Gamma(p,k)  &= \frac{1}{\varepsilon_{p}} \left \{ \Delta  E_k ({\bf k} \cdot {\bf p}) - 2 p^2 k^2 \right \}\\
&\quad + 2 \varepsilon_{p} ({\bf k} \cdot {\bf p}) \bigg(1 - \frac{\Delta}{2 \varepsilon_{k}} \bigg).\\
\end{split}
\end{equation}
In the above expression, we have used the following definitions
\begin{equation}\label{Ek}
E_k = \varepsilon_{k} - \Delta/2,  \  \  \varepsilon_{k} \equiv + \sqrt{k^2+[\Delta^2/4]}.
\end{equation}

Finally we derive the total transition moment as the sum of the bare (Eq.~(\ref{bareTMDC})) and the Coulomb interaction dependent spin-flip transition moment (Eq.~(\ref{dmTMDC})),
\begin{equation}\label{corrTMDC}
    \mu  + \delta \mu  =
    \langle \downarrow |2 S_x | \uparrow \rangle   ( 1 + \alpha (2\lambda/\Delta)^2 H(2k/\Delta) ),
\end{equation}
where the correction term is conveniently written as
\begin{equation}\label{delMTMDC}
\delta \mu = \langle \downarrow |2 S_x | \uparrow \rangle\alpha(2\lambda/\Delta)^2 H(2k/\Delta),
\end{equation}
with the function $H(2k/\Delta)$ which can be easily evaluated using Eq.~(\ref{dmTMDC}) \& Eq.~(\ref{gammaTMDC}). 

In the top panel of Fig.~\ref{fig:TMDC} we show the variation of the Coulomb interaction correction function $H(2k/\Delta)$ for $\alpha =4.095$ with respect to the dimensionless band momenta. We observe that the magnitude of this correction is very small for this class of materials with little-to-no variation. Thus the effect of Coulomb interaction correction is the least on the spin-flip transition moment in this class of materials. This can be understood from the fact that the spin-independent gap for these materials overwhelms the contribution from the spin-orbit coupling term. Hence this class of materials does not offer the unique tunability of the Coulomb-interaction dependent effect of the spin-flip transition moment, in a sense that the interaction  corrections are  negligibly small
for all reasonable values of   $\alpha$. Additionally, from Eq.~(\ref{alpha}), which takes into account effects beyond one-loop within the RPA self-consistency for the Kane-Mele model,  we concluded that  the Coulomb interaction correction effects  decreased further. Similar RPA calculations can be performed for this class of materials, however due to the intrinsic smallness of the 1-loop results in this case, the RPA formalism only leads to a small additional decrease of the already-small correction effect.

In the next section, we will derive the anomalous transition moment for the Silicene-Germanene class of materials. 

\section{Effect of Coulomb interactions on transition magnetic moments in Silicene-Germanene class of materials}
\label{sec:SG}
An application of  transverse electric field along the staggered sublattices of this class of materials causes the low-energy band structure to evolve from a topological insulator (TI) to a bulk insulator (BI) via a Valley Spin Polarized Metal (VSPM) state\cite{tabert1,ni,ezawa1,Ezawa2,falko}. In this section, we will first summarize the low-energy band structure of these class of materials and show that the evolution of the low-energy band structure from TI to BI via a VSPM state can also be attained with proper tuning of the dimensionless parameter which represents the ratio of spin-independent gap to the spin-orbit coupling ($\Delta/\lambda$). This class of materials thus open up the possibility to explore the Coulomb interaction correction for a large parameter regime. 

\begin{figure}[t]
\begin{center}
\includegraphics[width=\columnwidth]{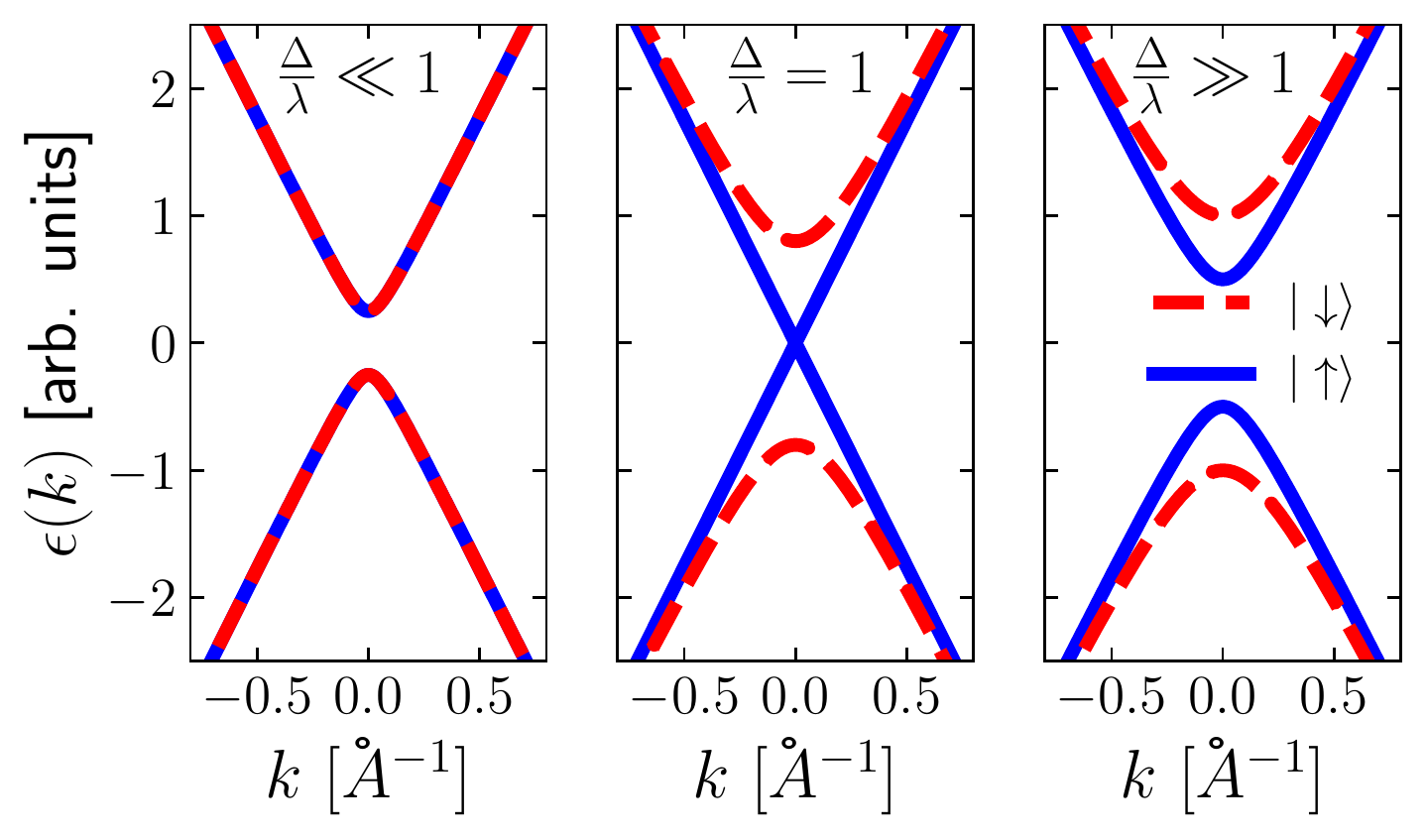} 
\caption{\label{fig:DCSG} For the purposes of illustration we plot the evolution  of energy dispersion curves for Silicene-Germanene type of materials. With the proper tuning of the ratio of spin-independent gap to spin-orbit coupling ($\Delta/\lambda)$, the band structure shows a transition from a topological insulator ($\Delta \ll \lambda$) to a band insulator ($\Delta \gg \lambda$) via the quantum critical VSPM state ($\Delta =\lambda$). Subsequent removal of the spin degeneracy is also observed for the bulk insulator regime ($\Delta \gg \lambda$). } 
\end{center}
\end{figure}

The Hamiltonian of this class of materials \cite{tabert1,ezawa1,Ezawa2} is
\begin{equation}\label{hamsili}
    H = v {\bf\sigma} \cdot {\bf k} + \frac{(\Delta - \lambda s_z)}{2} \sigma_z .
\end{equation}

The exact wave functions at momentum $k$ for the Hamiltonian given by Eq.~(\ref{hamsili}),
\begin{equation}\label{eigenfuncsili}
\Psi(k)_n^{s} = \frac{k}{\sqrt{k^2 + (E_{k,n}^s)^2}}
\left( \begin{array}{c} 1 \\
E_{k,n}^s/(k_x-ik_y)
\end{array}
\right) , \  \  n=1,2
\end{equation}
where $n=1$ labels the conduction band;
$n=2$ labels the valence band. We have defined $E_{k,n}^s$ as:

\begin{equation} \label{defE}
E_{k,n}^s = \varepsilon_{k,n}^s - \frac{(\Delta - \lambda s)}{2}, \  \  \   s=s_z =\pm 1,
\end{equation}
where the eigenenergies associated with the conduction and valence band are given by $\varepsilon_{k,n}^s$:
\begin{equation}\label{eigenenergysili}
\varepsilon_{k,1}^s  = \varepsilon_{k}^s \  > 0,  \  \ n=1,  
\end{equation}
\begin{equation}
\varepsilon_{k,2}^s  =   - \varepsilon_{k}^s  \ < 0,  \  \   n=2,
\end{equation}
and we use the definition:
\begin{equation}\label{defeigen}
\varepsilon_{k}^s \equiv + \sqrt{k^2+[(\Delta -\lambda s)^2/4]}.
\end{equation}

Fig.~\ref{fig:DCSG} shows the low-energy band structure corresponding to Eq.~(\ref{defeigen}) plotted for three different values of $(\Delta/\lambda)\ll 1$, $(\Delta/\lambda)= 1$ and $(\Delta/\lambda)\gg 1$. As can be seen the three different cases corresponding to different values of $(\Delta/\lambda)$ are consistent with the TI, VSPM and a BI state. General values of the spin-orbit coupling for these class of materials are in the range of $\lambda \approx 3\sim 40$ meV \cite{spinger,ezawaprl}. Next, we derive the expression for the bare transition moment using the conduction band wave functions given by Eq.~(\ref{eigenfuncsili}).

 From Eq.~(\ref{eigenfuncsili}), we can write the the corresponding conduction band ($n =1$) wave functions $|\uparrow\rangle$ and $|\downarrow\rangle$ as:
\begin{equation}\label{cup}
|\uparrow\rangle= \Psi(k)_1^{+} = \frac{k}{\sqrt{k^2 + (E_{k,1}^+)^2}}
\left( \begin{array}{c} 1 \\
E_{k,1}^+/(k_x-ik_y)
\end{array}
\right)
\end{equation}
\begin{equation}\label{cdown}
|\downarrow\rangle = \Psi(k)_1^{-} = \frac{k}{\sqrt{k^2 + (E_{k,1}^-)^2}}
\left( \begin{array}{c} 1 \\
E_{k,1}^-/(k_x-ik_y)
\end{array}
\right)
\end{equation}
Using Eq.~(\ref{cup}) and Eq.~(\ref{cdown}), we derive an expression for the band-momentum dependent bare transition moment,
\begin{equation}\label{baresili}
\mu = 2\langle \downarrow | S_x | \uparrow \rangle = \frac{k^2 + E_{k,1}^{+}E_{k,1}^{-}}{\sqrt{(k^2 + [E_{k,1}^{+}]^2)(k^2 + [E_{k,1}^{-}]^2)}}.
\end{equation}

\begin{figure}[t]
\begin{center}
\includegraphics[width=\columnwidth]{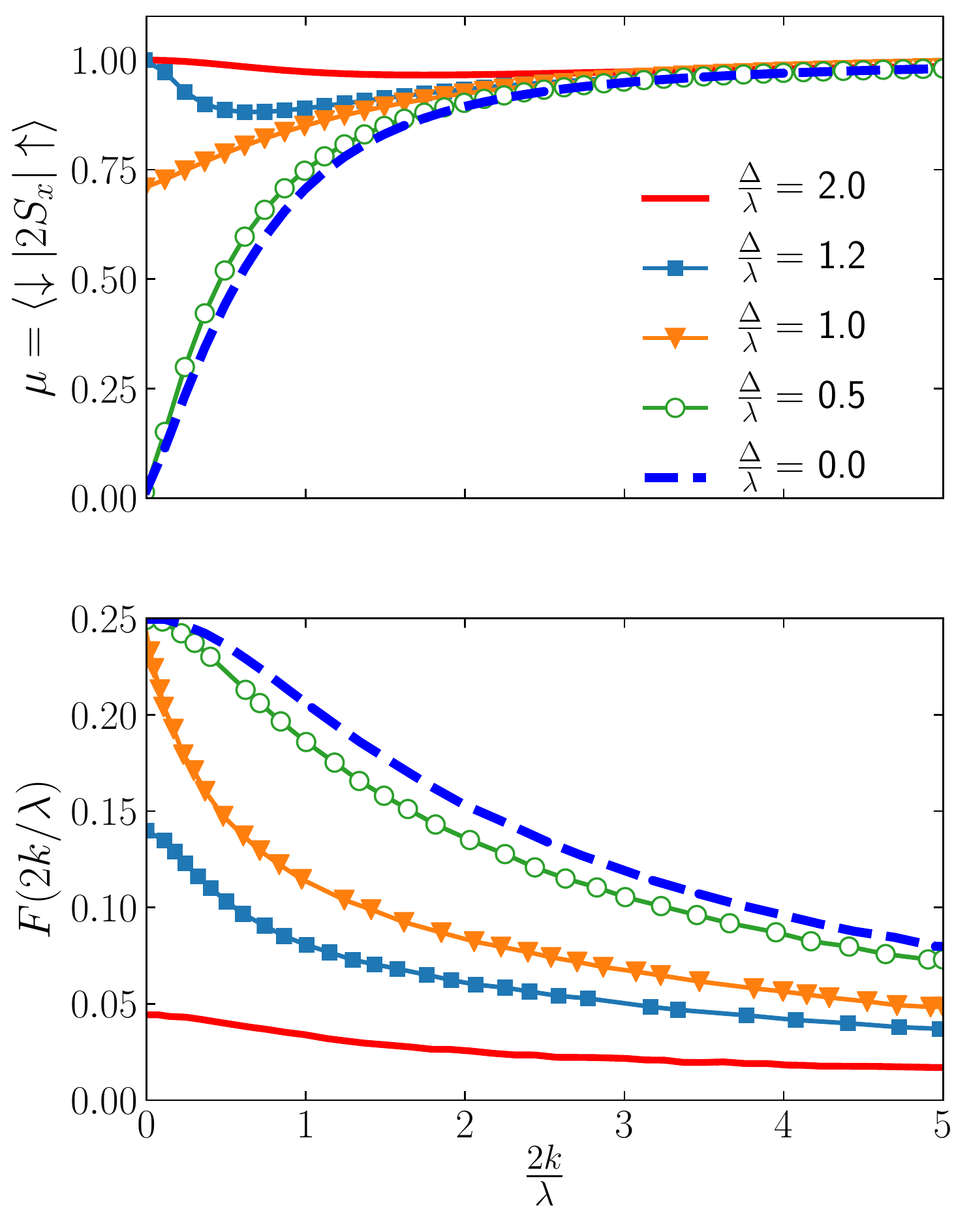} 
\caption{\label{fig:SGF} Top panel: Variation of the bare transition moment $\mu = \langle\uparrow|2S_{x}|\downarrow\rangle$ with the rescaled momentum $2k/\lambda$ for several values of $\Delta/\lambda$. As the coupling parameter $\Delta/\lambda$ increases the system makes a transition from TI to BI via the VSPM state ($\Delta/\lambda =1$). Bottom panel: Variation of the correction function $F(2k/\lambda)$ with $2k/\lambda$ for various values of coupling $0 <  \Delta/\lambda < 2 $. The correction term is seen to be large for the topological insulators ($\Delta/\lambda \ll 1$) compared to the VSPM ($\Delta/\lambda =1$) or bulk insulators states ($\Delta/\lambda \gg 1$).}
\end{center}
\end{figure}


In the top panel of Fig.~\ref{fig:SGF}, we show the variation of the bare transition moment  $\mu\equiv \langle \downarrow | 2S_x | \uparrow \rangle$ with a dimensionless rescaled momentum $(2k/\lambda)$ for various values of $(\Delta/\lambda)$. 
The overall variation of the spin flip transition with momentum for different values of the spin-independent gap can be intuitively understood in the following way.
At $\Delta/\lambda=0$, which corresponds to the Kane Mele model, the system is stiff in the spin $z$ direction as the term proportional to $s_z$ favors ordering and thus
a transverse field at zero momentum (uniform field) can not cause spin flip, while at  finite band momentum this becomes possible due to presence of the kinetic energy term. 
 In the opposite extreme, $\Delta/\lambda \gg 1$, a spin flip can be achieved effortlessly as the
 $s_z$ term can be neglected. 

Finally we turn to the interaction corrections. 
The expression for the Green's function G({\bf k},$\omega$) corresponding to the Hamiltonian of Eq.~(\ref{hamsili}), reads
\begin{equation}\label{GFsili}
\begin{split}
    G({\bf k},\omega)& = \frac{1}{2\varepsilon_{k}^{s}}\bigg[\frac{\varepsilon_{k}^{s}+{\bf\sigma} \cdot {\bf k}+\big(\frac{\Delta-\lambda s_{z}}{2}\big)\sigma_{z}}{(\omega-\varepsilon_{k}^{s} +i \eta)}\\
    &\quad +\frac{\varepsilon_{k}^{s}-{\bf\sigma} \cdot {\bf k}-\big(\frac{\Delta-\lambda s_{z}}{2}\big)\sigma_{z}}{(\omega+\varepsilon_{k}^{s} -i \eta)}\bigg].\\
\end{split}
\end{equation}

We evaluate the frequency integral in Eq.~(\ref{ICsiliKM}) by using the Green's function expression from Eq.~(\ref{GFsili}) resulting in 
\begin{equation}\label{FIsili}
\begin{split}
    i\int\frac{\mathrm{d}\omega}{2\pi}\bigg[G^{-}G^{+}\bigg]&=\frac{1}{2\varepsilon_{p}^{-}\varepsilon_{p}^{+}}\bigg\{\frac{1}{(\varepsilon_{p}^{+} + \varepsilon_{p}^{-})}\bigg\}\bigg[\varepsilon_{p}^{-}\varepsilon_{p}^{+} - ({\bf \sigma} \cdot {\bf p})^{2} \\
    &\quad +\bigg(\frac{\lambda^{2}-\Delta^{2}}{4}\bigg) + \lambda ({\bf \sigma}\cdot {\bf p})\sigma_{z}\bigg].\\
\end{split}
\end{equation}
We substitute the above equation along with the conduction band wave functions given by Eqs.~(\ref{cup}) and (\ref{cdown}) in Eq.~(\ref{ICsiliKM}) to derive the expression for the correction term:
\begin{equation}\label{delmsili}
\begin{split}
    \delta \mu &=
    \langle \downarrow |2 S_x | \uparrow \rangle \sum_{{\bf p}} \frac{V(|{\bf p}-{\bf k}|)}{2\varepsilon_{p}^{+}\varepsilon_{p}^{-}
    (\varepsilon_{p}^{+}+\varepsilon_{p}^{-})}\bigg[ \varepsilon_{p}^{+}\varepsilon_{p}^{-} - p^2\\  &\quad +\bigg(\frac{\lambda^2-\Delta^2}{4}\bigg) + \lambda  ({\bf p}\cdot {\bf k}) \bigg\{\frac{\varepsilon_{k}^{-}-\varepsilon_{k}^{+} - \lambda}{k^2 + E_{k,1}^{+}E_{k,1}^{-}} \bigg\}\bigg]\\
    &\quad = \langle \downarrow | 2S_x | \uparrow \rangle\alpha F(2k/\lambda,\Delta/\lambda).
\end{split}
\end{equation}

The function $F(2k/\lambda, \Delta/\lambda)$ quantifies the Coulomb interaction correction effects and we have defined it as:
\begin{equation}\label{Fsili}
\begin{split}
    \alpha F(2k/\lambda,\Delta/\lambda) & \equiv  \sum_{{\bf p}} \frac{V(|{\bf p}-{\bf k}|)}{2\varepsilon_{p}^{+}\varepsilon_{p}^{-}
    (\varepsilon_{p}^{+}+\varepsilon_{p}^{-})} \bigg[ \varepsilon_{p}^{+}\varepsilon_{p}^{-} - p^2\\
    &\quad + \bigg(\frac{\lambda^2-\Delta^2}{4}\bigg)+ \lambda  ({\bf p}\cdot {\bf k})\\
    &\quad \times \bigg\{\frac{\varepsilon_{k}^{-}-\varepsilon_{k}^{+} - \lambda}{k^2 + E_{k,1}^{+}E_{k,1}^{-}}\bigg\} \bigg],
\end{split}
\end{equation}
with $\alpha= e^{2}/ \epsilon \hbar v$ as the effective fine-structure constant that gives the strength of the interactions.

Using Eq.~(\ref{baresili}) and Eq.~(\ref{delmsili}), we write the expression for the total spin-flip transition moment as
\begin{equation}\label{Total moment}
\mu + \delta \mu = \langle \downarrow | 2S_x | \uparrow \rangle \bigg\{ 1 + \alpha F(2k/\lambda, \Delta/\lambda) \bigg\}.
\end{equation} 
To assess quantitatively  the effect of the Coulomb interaction as a function of the band momentum, we plot the variation of the function $F(2k/\lambda)$ with the rescaled momentum ($2k/\lambda$) for several values of ($\Delta/\lambda$) in the bottom panel of Fig.~\ref{fig:SGF}. The correction function is seen to be maximum at $k=0$ for all the different values of $(\Delta/\lambda)$ and is seen to decrease with increasing values of the band momenta. Although for $(\Delta/\lambda)\leq 1$ the bare transition moment $\mu$ was found to be 0 at $k=0$, the Coulomb interaction correction effects turn out to be the largest for this regime. However increasing the value of ($\Delta/\lambda$) leads to a decrease in the Coulomb interaction correction. Of course the correction function $F(2k/\lambda)$   has to be multiplied by the dimensionless Coulomb interaction strength $\alpha \sim 1$ which is strongly material and environment dependent. It is clear from Fig.~\ref{fig:SGF} that the overall interaction effect is strongest in the  parameter regime $\Delta/\lambda \approx 0$, i.e. in the Kane Mele universality class, while  for $\Delta/\lambda  > 1$ and beyond the correction becomes gradually smaller and less pronounced even for substantial values of $\alpha$ as the system becomes dominated by the spin-independent gap.

\section{Discussion and Outlook}
\label{sec:sum}
In summary we have analyzed, for the first time, the  effect of Coulomb interactions on the spin transition magnetic moment for the case of atomically thin hexagonal lattices with spin-orbit interactions, such as 2D topological insulators (described by the Kane Mele Model), dielectric group-VI Dichalcogenides and the Silicene-Germanene class of materials. Due to the non-relativistic nature of these systems,  and  because of the two-dimensional nature of all the studied materials (meaning that the spin-orbit
interaction is a relatively small effect on top of the band structure),  the ``anomalous", i.e. Coulomb interaction effect manifests itself anisotropically, and indeed only in the spin-flip channel, for magnetic fields in the material planes. This is in contrast (although conceptually and technically very similar in spirit) to the famous anomalous magnetic moment  of  the electron  in relativistic QED where
the Schwinger result renormalizes directly and isotropically the electron g-factor.
We can view our results as yet another important manifestation of  (moderately strong)  electron-electron interaction effects 
in graphene-like hexagonal monolayer systems which exhibit Dirac quasiparticle spectra. 

As discussed in the previous section which contains results across all parameter regimes (Fig.~\ref{fig:SGF}), it appears  that  the Kane Mele limit (i.e. no spin-independent gap, but a gap induced by the spin-orbit interaction) represents the point in parameter space where the Coulomb corrections are the strongest (Fig.~\ref{fig:DCKM}).
On the other hand, the monolayer dichalcogenides which are characterized by $\alpha$ as large as $\alpha \approx 4 $ (much larger than suspended graphene with SOC) have relatively large gaps, but reside firmly in the parameter regime $\Delta/\lambda  > 1$ making the anomalous effects  much smaller and therefore harder to detect (see the relevant Fig.~\ref{fig:TMDC}).
 We also note that our calculations were performed to first order in the bare Coulomb interaction $\alpha$ when the interaction effects are small, while we have used the RPA  approximation, which takes into account self-consistent screening, for large bare $\alpha$ (relevant to suspended samples). The difference between  the two approaches  is important in practice only for the Kane-Mele model. 
 Additionally, our work displays that the control of interactions can be achieved for example
by using different substrates which can affect the Coulomb interaction via different levels of dielectric screening. 


The anomalous spin contributions investigated in this work could lead to detectable signatures in experiments sensitive to  spin relaxation/decoherence phenomena. For the case of sufficient spin-orbit coupling and band gaps in the  range of $\sim$ 1 eV, a very promising  magneto-optical Kerr effect technique previously employed to measure spin decoherence times \cite{crooker,furis} may be sensitive enough to detect such anomalous contributions. However, it is important to emphasize that  the spin relaxation mechanism in 2D materials is very material-specific and
 depends strongly  on various parameters such as  ripples, phonons, nature of substrates and magnetic impurities \cite{hernando,ertler, fratini, lundeberg,Tuan2014,castroneto09}. It would be  interesting to investigate the effect of anomalous spin contributions on the spin relaxation mechanism with the inclusion of various dissipative effects, such as phonons, ripples and impurities. A microscopic theory that studies the effect of anomalous spin contributions on spin-flip lifetimes is well beyond the scope of the present work and is left for the future.





\section*{acknowledgments}
SS is grateful to Prof. Ion Garate for insightful discussion during the initial formulation of this work. SS was funded by the Canada First Research Excellence Fund. 
VNK gratefully acknowledges the financial support of the Gordon Godfrey visitors program at the School of Physics, University of New South Wales,  Sydney,
during two research visits. VNK also acknowledges financial support from NASA grant number 80NSSC19M0143.


\bibliographystyle{apsrev4-1}
\bibliography{SSGrefs_Valeri}
\end{document}